\documentstyle[aps,multicol,psfig]{revtex}

\begin{document}
\draft
\title{Effective density dependent pairing forces in the T=1 and T=0 channels.}
\author{E. Garrido$^{1}$, P. Sarriguren$^{1}$, E. Moya de Guerra$^{1}$,
and P. Schuck$^{2}$}
\address{$^{1}$ Instituto de Estructura de la Materia, Consejo Superior de\\
Investigaciones Cient\'{\i }ficas, Serrano 123, E-28006 Madrid, Spain\\
$^{2}$ Institut des Sciences Nucl\'{e}aires, Universit\'{e} Joseph Fourier,\\
CNRS-In2P3\\
53, Avenue des Martyrs, F-38026 Grenoble Cedex, France}
\date{\today }
\maketitle

\begin{abstract}
Effective density dependent pairing forces of zero range are adjusted on gap
values in $T=0,1$ channels calculated with the Paris force in symmetric
nuclear matter. General discussions on the pairing force are presented. In
conjunction with the effective k-mass the nuclear pairing force seems to
need very little renormalization in the $T=1$ channel. The situation in
the $T=0$ channel is also discussed.
\end{abstract}

\pacs{21.65.+f; 21.30.+y; 21.10.-k; 05.30.Fk}

\begin{multicols}{2}

\section{Introduction}

The novel availability of exotic nuclei has spurred an immense revival of
nuclear structure investigations \cite{nupecc}. Indeed nuclei close to the
neutron or proton drip lines may exhibit very unusual features such as
pronounced neutron or proton skins \cite{fuku}, and neutron halos 
\cite{hansen}. Among many very interesting questions, nuclear pairing has 
again become on the forefront of theoretical interest. Indeed the
existence of
neutron halos is due to the pairing force \cite{BE,halos} and in heavier
proton rich $N\simeq Z$ nuclei the proton-neutron pairing may play an
important role \cite{nz}. In this work we therefore want to address some
problems of neutron-neutron and proton-neutron pairing. This concerns for
instance considerations of the effective pairing interactions. However, we
also will discuss some other aspects of more general character. We mostly
will study the infinite matter case.

\section{Generalities on the nuclear pairing forces}

It is a well established fact that, aside from the exception of magicity,
nuclei are superfluid. There can be $nn$ as well as $pp$ pairing whereas $pn$
pairing is less frequent. One of the main questions we will treat here is
the effective pairing force. We will do this in the framework of homogeneous
nuclear matter at various densities. The limit to finite nuclei can be
established through the Local Density Approximation (LDA) which seems to
work very well also for the nuclear pairing problem \cite{kucha}. Quite
generally the equation for the gap $\Delta $ in nuclear matter can be
written as

\begin{equation}
\Delta _{{\bf p}}=-\sum_{{\bf k}}v_{{\bf pk}}\;\frac{\Delta _{{\bf k}}}
{2\sqrt{\left( \epsilon _{k}-\epsilon _{F}\right) ^{2}+
\Delta _{{\bf k}}^{2}}}
\label{gapeq}
\end{equation}
where $v_{{\bf pk}}$ is the (effective) pairing force, the $\epsilon _{k}$
are the Brueckner-Hartree-Fock single particle energies and $\epsilon _{F}$
is the Fermi energy. The summation goes over momentum states. In 
(\ref{gapeq}) we did not specify whether we consider the $T=1$ or $T=0$ 
channels.

The first aspect we want to discuss is what kind of force $v_{pk}$ shall be
used in Eq.(\ref{gapeq}) from a microscopic point of view. The answer to
this question is in principle very well known since the early days of
superconductivity and superfluidity. Since the gap equation can be derived
from the Bethe-Salpeter equation for the two particle many-body Green's
function \cite{RS,migdal}, the pairing force $v_{pk}$ is built out of the
sum of all particle-particle irreducible Feynman graphs \cite{RS,migdal}. To
lowest order in the bare interaction it is given by Fig. 1.
\end{multicols}

\begin{figure}[ht]
\centerline{\psfig{figure=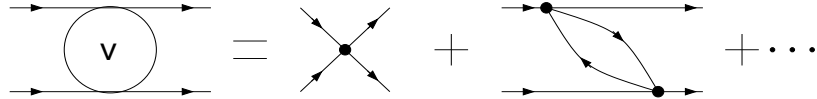,width=12.5cm,%
bbllx=9.0cm,bblly=-1.5cm,bburx=12.7cm,bbury=25.2cm,angle=270}}
\vspace{0.2cm}
\caption[]{
Schematic representation of the pairing force $v_{pk}$ to
lowest order in the bare interaction.
}
\end{figure}

\begin{multicols}{2}

In Fig. 1 the dot stands for the bare vertex. The second term represents a
$ph$ screening correction to the bare force. The very important point we want
to make here is that in no way some type of Bethe-Goldstone or Brueckner
G-matrix can be used in the gap equation. Since the gap equation is already
a kind of in medium two-body Schroedinger equation (see e.g. Ref. \cite
{henl,BLS}) one cannot use a G-matrix which in itself is a solution of the
in medium two-body problem. Otherwise there is severe double counting. Since
the G-matrix essentially softens the short range repulsion one expects that
pairing becomes enhanced if used in the gap equation. In the pairing problem
everything depends exponentially on the system parameters \cite{expon} and
this effect can then be quite large. A demonstration is given in Fig. 2,
where the $nn$ gap is calculated once with the bare Paris force \cite{paris}
and once with the corresponding G-matrix \cite{baldo90}. One sees that the
use of the G-matrix enhances the gap value by practically a factor of two.

\begin{figure}[t]
\centerline{\psfig{figure=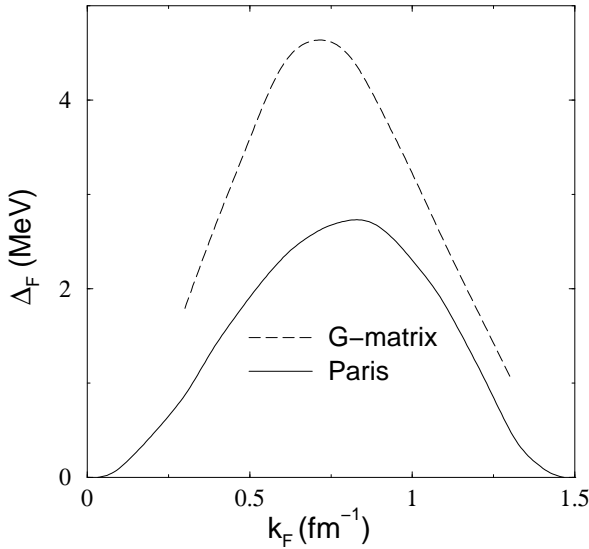,width=12.5cm,%
bbllx=3.0cm,bblly=-3cm,bburx=20cm,bbury=25.2cm,angle=270}}
\vspace{0.2cm}
\caption[]{
Pairing gap $\Delta _{F}$ in neutron matter as a function of
the Fermi momentum $k_{F}$ with the Paris force and with the corresponding
G-matrix.
}
\end{figure}

Sometimes Eq.(\ref{gapeq}) is divided into a low momentum and a high
momentum space and the high momentum space is eliminated in renormalizing
consistently the bare interaction in the low momentum space \cite{baldo90}.
This type of procedure is, of course, perfectly allowed, since it is only a
mathematical trick for solving Eq.(\ref{gapeq}). Unfortunately in nuclear
physics it is a quite widespread habit since decades (see for example 
\cite{habit} and the critics given in \cite{henl}) to use some kind of 
G-matrix
in Eq.(\ref{gapeq}) as for example Skyrme forces which are to be considered
as a phenomenological representation of a microscopic G-matrix. One will
object that one of the most successful nuclear $nn$ pairing forces, namely
the Gogny force \cite{gogny} is also to be considered as a G-matrix. Things
are, however, more subtle there as we now will explain. The first
observation is that the Gogny force in the $^{1}S_{0}$ channel is of finite
range but density independent. Second, one finds when solving the gap
equation with the Gogny force in nuclear matter that it gives results which
are very close to the ones obtained with the Paris force or any other
realistic bare nucleon-nucleon force. This is demonstrated in Fig. 3 where
we compare results of the gap from the D1, D1S, and Paris forces.

\begin{figure}[t]
\centerline{\psfig{figure=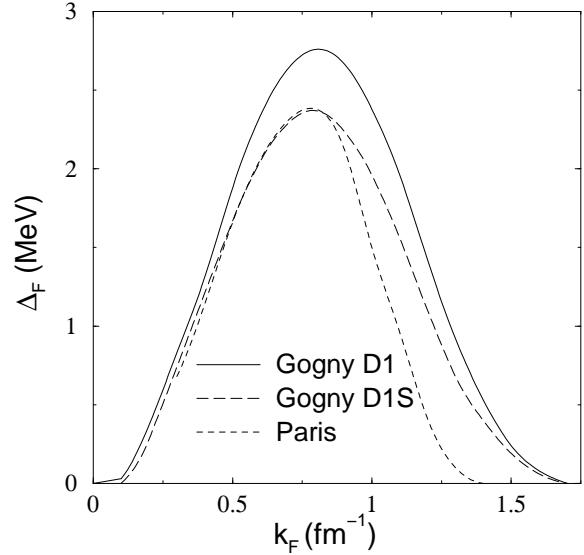,width=12.5cm,%
bbllx=3.0cm,bblly=-3cm,bburx=20cm,bbury=25.2cm,angle=270}}
\vspace{0.2cm}
\caption[]{
Pairing gap $\Delta _{F}$ in the $^{1}S_{0}$ channel in
symmetric nuclear matter calculated with the Gogny forces D1 and D1S,
compared with the Paris force results, from Refs. \protect\cite{baldo90,kucha}
}
\end{figure}

We see that D1S is still much closer to Paris than D1. Indeed D1S has been
readjusted \cite{d1s} to give in first place a lower surface tension than D1
but at the same time to give a smaller even-odd staggering so that it
becomes in closer agreement with experiment. It is very surprising that this
readjustment brought D1S so close to the bare Paris force. So in the 
$S=0\;T=1$ channel the Gogny force acts like a realistic bare force at least
in what concerns energies up to the Fermi energy. This conclusion was also
found in \cite{BE} and is further confirmed by the fact that the scattering
length corresponding to D1S, $a_{D1S}=12.16$ fm, is very large and of the
same order of magnitude as the experimental value $a=18.5$ fm.

The reason why the Gogny force acts like a free force in the $nn$ pairing
channel in spite of the fact that it has been adjusted to the G-matrix from
the Sprung-Tourreil force \cite{sprung} can only be guessed: probably for
this force in that channel the Pauli blocking is so efficient that in the
G-matrix equation, $G=v+v\frac{Q}{e}G$ , the second term on the r.h.s. is
suppressed. On the other hand the question remains why experiment apparently
demands a pairing force very close to the bare one. This is true at least in
the $T=1$ channel. For the $T=0$ channel much less investigations have been
performed and it is unclear whether a bare force can be used as well. One
reason which can be advanced to explain the validity of the bare force is a
possible cancellation between screening effects and effective mass
enhancement. Graphically these two possibly opposing effects are shown to
lowest order in the interaction in Fig. 4.

\end{multicols}

\begin{figure}[ht]
\centerline{\psfig{figure=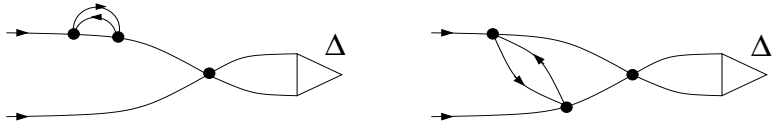,width=12.5cm,%
bbllx=9.0cm,bblly=-1.5cm,bburx=12.7cm,bbury=25.2cm,angle=270}}
\vspace{0.2cm}
\caption[]{
Schematic representation of the effective mass enhancement
and of the screening effect to lowest order in the interaction.
}
\end{figure}

\begin{multicols}{2}

In this respect it should be mentioned that the Hartree-Fock-Bogoliubov
(HFB) calculations with the Gogny force are performed with the so called 
$k-$mass $m^{*}<m$. However one knows that the corresponding level density
close to the Fermi energy is much too small. Including $E-$mass
corrections such
as the one shown in Fig. 4 brings the effective mass at the Fermi level back
to the bare mass or even overshoots it. For consistency the screening of the
bare force also shown in Fig. 4 must be included. Larger effective masses
enhance pairing while screening probably weakens it so that the net effect
could be the bare force. To investigate such effects, extreme care must be
taken that both contributions of Fig. 4 are treated on the same footing.
Since, as already mentioned, pairing depends exponentially on the system
parameters, the slightest imbalance (for example in treating both graphs of
Fig. 4 in slightly different approximations) may cause strong erroneous
results. One way to treat things consistently could be to use the Gorkov
equations \cite{gorkov} and develop the normal and abnormal parts of the
mass operator matrix to second order Born approximation and solve the
corresponding gap equation numerically. In medium effects similar to the
ones shown in Fig. 4 have been included in the past to the pairing problem
in one way or the other \cite{wam}. Practically all calculations resulted in
an important reduction of $\Delta =\Delta \left( k_{F}\right) $ compared to
the values shown in Fig. 3. It can be deduced from the study in \cite{kucha}
that a reduction of pairing in infinite matter obtained with the Gogny force
in a global way, i.e. for all values $0\leq k_{F}\leq 1.4$ fm$^{-1}$,
inevitably leads also to a reduction of pairing in finite nuclei of the same
proportions (this fact can be understood via the local density approximation
which as mentioned already, on average, yields comparable results to quantal
calculations \cite{kucha,beng}). It therefore can be concluded that e.g. a
reduction of the $\Delta -$values in Fig. 3 by a factor of two (a scenario
often encountered in the calculations of references mentioned above) will
fail to reproduce experimental gap values of nuclei when the underlying
theory is applied to finite nuclei.

Concluding these general considerations we want to say that in absence of
any necessity stemming from experimental facts it is probably safe to treat
nuclear pairing in conventional mean field theory with the bare
nucleon-nucleon potential as this is indicated from the microscopic theory
and as apparently is needed to reproduce experimental facts in the $T=1$
channel. Using this philosophy one arrives naturally for $T=0\;np\;$ pairing
at much stronger gap values \cite{BLS} since the $N-N$ force is strongest in
this channel. We will give some more details about this in the next section
and also discuss how the bare interaction in the gap equation can be
replaced by an equivalent density dependent zero range force such as they
have become quite popular recently in the nuclear structure problem.

\section{Effective density dependent zero range pairing forces}

In the last section we gave arguments that, at least as a first guess, it is
indicated to use as the pairing force the bare nucleon-nucleon potential. We
here want to develop arguments that this strategy is not necessarily
orthogonal to the popular employment of density dependent zero range forces
with a cutoff. Such arguments have first been developed by Bertsch and
Esbensen \cite{BE} and we here want to refine these arguments on the one
hand and on the other side extend them also to $T=0\;np$ pairing.

\begin{figure}[t]
\centerline{\psfig{figure=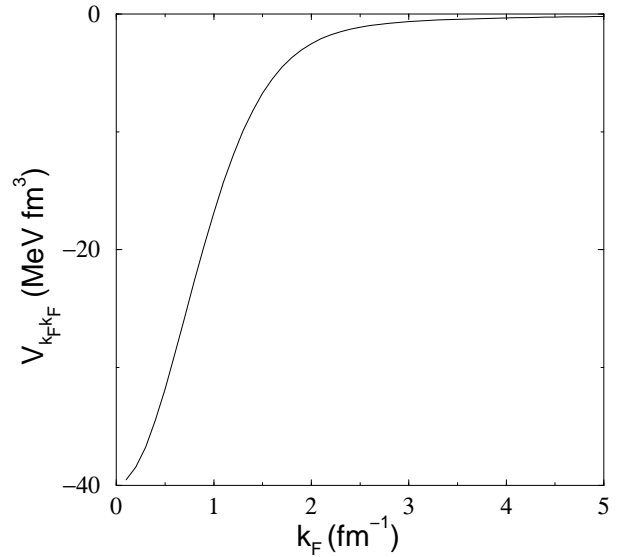,width=12.5cm,%
bbllx=3.0cm,bblly=-3cm,bburx=20cm,bbury=25.2cm,angle=270}}
\vspace{0.2cm}
\caption[]{
$v_{k_{F}k_{F}}$ vs. $k_{F}$ in the $S=0\;T=1$ channel for
the Gogny D1 force.
}
\end{figure}

A qualitative argument why a density independent finite range force in the
gap equation (Eq.(\ref{gapeq})) can be replaced by a density dependent zero
range one with a cutoff, goes as follows. For $s-$wave pairing only the
angle averaged matrix element $\tilde{v}_{pk}$ enters the gap equation 
$\Delta _{p}=\sum_{k}\tilde{v}_{pk}\kappa _{k}$, where $\kappa _{k}=\Delta
_{k}/2E_{k}$ is the abnormal density and 
\begin{equation}
E_{k}=\sqrt{\left( \epsilon _{k}-\epsilon _{F}\right) ^{2}+\Delta _{k}^{2}}
\label{qpe}
\end{equation}
the quasiparticle energy. The former is very much peaked at $k=k_{F}$ with a
peak width of the order $\Delta =\Delta _{k_{F}}$. Since anyway in pairing
problems only the gap values at $k\simeq k_{F}$ matters, we see that for 
$\Delta _{k_{F}}$ only the value of the matrix element 
$\tilde{v}_{k_{F}k_{F}} $ plays a significant role. In the Gogny force, 
this matrix
element as a function of $k_{F}$ is shown in Fig. 5. Since a $\delta -$force
is a constant in $k-$space, one has to weight the $\delta -$force with a 
$k_{F}$, i.e., a density dependent factor similar to $\tilde{v}_{k_{F}k_{F}}$
in order to recover the essential pairing features of the original finite
range force. The only thing we have to add is a cutoff value, otherwise the
gap equation would diverge. Bertsch and Esbensen \cite{BE} therefore
proposed the following density dependent zero range force

\begin{equation}
v\left( {\bf r}_{1},{\bf r}_{2}\right) =v_{0}\left[ 1-\eta \left( \frac{\rho
\left( \frac{{\bf r}_{1}+{\bf r}_{2}}{2}\right) }{\rho _{0}}\right) ^{\alpha
}\right] \delta \left( {\bf r}_{1}-{\bf r}_{2}\right)  \label{bert}
\end{equation}
where $v_{0},\eta ,\alpha $ are adjustable parameters and $\rho _{0}$ is the
saturation density. In the gap equation (\ref{gapeq}), Eq.(\ref{bert}) must
be supplemented with a cutoff value $\epsilon _{C\text{ }}$ which thus
constitutes a fourth parameter. However, at zero density the cutoff and 
$v_{0}$ must be chosen such that the scattering length $a$ is reproduced. 
For Eq. (\ref{bert}) one obtains the relation

\begin{equation}
v_{0}=-\frac{\hbar ^{2}}{m}2\pi ^{2}\frac{1}{\frac{\pi }{a}+k_{C}}
\label{v0}
\end{equation}
where $\frac{k_{C}^{2}}{2m}=\epsilon _{C}$. The neutron-neutron scattering
length is very large $\left( a=18.5\text{ fm}\right) $ and we take in 
Eq.(\ref{v0}) the limit $a\rightarrow \infty $ that leads to the relation 
for $v_{0}$ also used in \cite{BE}. One finally remains with three adjustable
parameters $\left( \eta ,\alpha ,\epsilon _{C}\right) $and the gap equation
reads

\begin{eqnarray}
\lefteqn{
1=}    \label{eqber} \\ &&
-\frac{v_{0}}{\pi ^{2}}\left[ 1-\eta \left( \frac{\rho }{\rho _{0}}\right)
^{\alpha }\right] \left( \frac{m^{*}\left( \rho \right) }
{2\hbar ^{2}}\right) ^{3/2}
\!\!\!\! \int_{0}^{\epsilon _{C}}\!\!\!\!\!\!\!d\epsilon 
\sqrt{\frac{\epsilon }{\left( \epsilon -\epsilon _{F}\right) ^{2}
+\Delta ^{2}}}  \nonumber
\end{eqnarray}
With respect to \cite{BE} we considered also a density dependent effective
mass. Since finite nuclei calculations are performed with such an effective
mass one must account for it when adjusting a $\delta -$force which later
shall be used in BCS or HFB calculations. For the effective mass we take the
one corresponding to the Gogny force,

\begin{eqnarray}
\left( \frac{m^{*}\left( \rho \right) }{m}\right) ^{-1}&
\!\!\!\!\!\!\!\!=&1+\frac{m}{2\hbar
^{2}}\frac{k_{F}}{\sqrt{\pi }}\sum_{c=1}^{2}\left[ W_{c}+2\left(
B_{c}-H_{c}\right) -4M_{c}\right] \nonumber \\
& & \mu _{c}^{3}e^{-x_{c}}\left[ \frac{\cosh
\left( x_{c}\right) }{x_{c}}-\frac{\sinh \left( x_{c}\right) }
{x_{c}^{2}}\right]  \label{meff}
\end{eqnarray}
with $x_{c}=k_{F}^{2}\mu _{c}^{2}/2,$ and the coefficients 
$W_{c},B_{c},H_{c},M_{c},\mu _{c}$ corresponding to the Gogny force 
D1\cite{RS,gogny}.

\begin{figure}[t]
\centerline{\psfig{figure=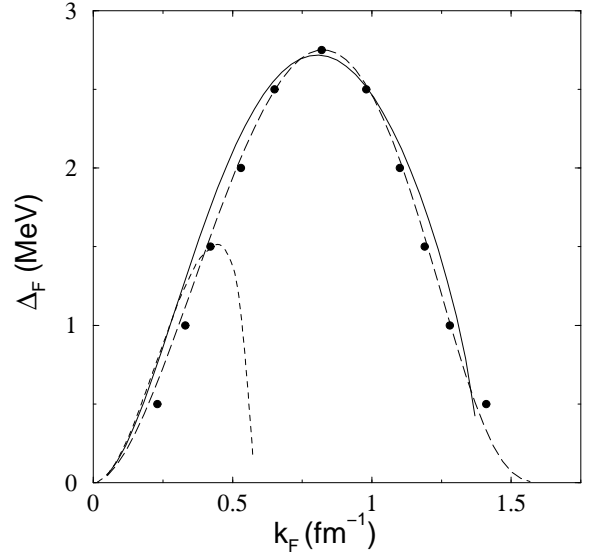,width=12.5cm,%
bbllx=3.0cm,bblly=-3cm,bburx=20cm,bbury=25.2cm,angle=270}}
\vspace{0.2cm}
\caption[]{
$T=1$ pairing gap in nuclear matter. The dots are the results
of a Hartree-Fock calculation using the Gogny force. The continuous (a) and
dashed (b) curves are the results obtained with the effective pairing
interaction in Eq.(\ref{eqber}) with (a) effective mass $m^{*}/m$ as in 
Eq.(\ref{meff}) and (b) $m^{*}/m=1$ (see text). The dotted line
corresponds to the pairing force in Ref. \cite{bonche} (see text).
}
\end{figure}

In Fig. 6 we show the fit to $\Delta \left( k_{F}\right) $ in the isovector
channel obtained from Eq. (\ref{eqber}) with $\epsilon _{C}=60$ MeV, $\eta
=0.45,\;\alpha =0.47$ . Also shown is the fit corresponding to the bare mass
(i.e., $m^{*}/m=1$) with $\epsilon _{C}=60$ MeV, $\eta =0.70,\;\alpha =0.45$
, as in Ref. \cite{BE}. In both cases, the corresponding $v_{0}$ value is 
$v_{0}=481$ MeV fm$^{3}$. We see that the fits are good for values of 
$k_{F}$ up to the saturation value $k_{F}=1.35$ fm$^{-1}.$ A density 
dependent $\delta -$force has also been used for $T=1$ pairing in finite 
nuclei in the
context of the HFB \cite{bonche} and in the context of relativistic
Hartree-Bogoliubov \cite{meng}. The strength used there is however larger.
If we use the pairing force in Ref. \cite{bonche} with $V_{0}=700$ MeV 
fm$^{3}$, we get the dotted line curve shown in Fig. 6 that corresponds to the
following parameters in our notation: $v_{0}=1400$ MeV fm$^{3}$; $\epsilon
_{C}=7$ MeV; $\eta =1$ MeV and $\alpha =1$ MeV.

For finite nuclei, the force (\ref{bert}) can be used in BCS approximation

\begin{equation}
\Delta _{i}=-\sum_{k,\epsilon _{k}\leq \epsilon _{C}}\left\langle 
i\bar{i}\left| v\right| k\bar{k}\right\rangle \frac{\Delta _{k}}{2E_{k}}  
\label{bcs}
\end{equation}
or in the HFB approach where the gap equation has the form (\ref{bcs}) in
the canonical basis. We want to point out that the cutoff has to be counted
relative to the bottom of the single-particle well and not from its edge.

\section{Proton-neutron pairing in the T=0 channel}

In this section we want to extend our considerations to $n-p$ pairing in the 
$T=0$, i.e. in the deuteron channel. As we suggested earlier, as a first
guess one should investigate the gap equation with the bare force. The gap
equation in homogeneous symmetric nuclear matter has recently been solved
for the $T=0$ channel \cite{BLS} using the bare Paris force with single
particle energies obtained in Brueckner-Hartree-Fock approximation. Since in
the deuteron channel $\left( T=0,S=1,L=0,2\right) $ we have a mixture of $s-$
and $d-$waves involving the tensor force, the net outcome is more attraction
leading to the deuteron bound state in free space. This increased attraction
then takes over to the gap equation (which in the zero density limit turns
into the Schroedinger equation for the deuteron, see \cite{BLS,noz}) and,
not unexpectedly (remember the exponential dependence), the gap values in
the $T=0$ channel as a function of $k_{F}$ are much stronger reaching values
more than a factor of two larger than in the $T=1$ channel. This is shown in
Fig. 7 (Ref. \cite{BLS}).

\begin{figure}[t]
\centerline{\psfig{figure=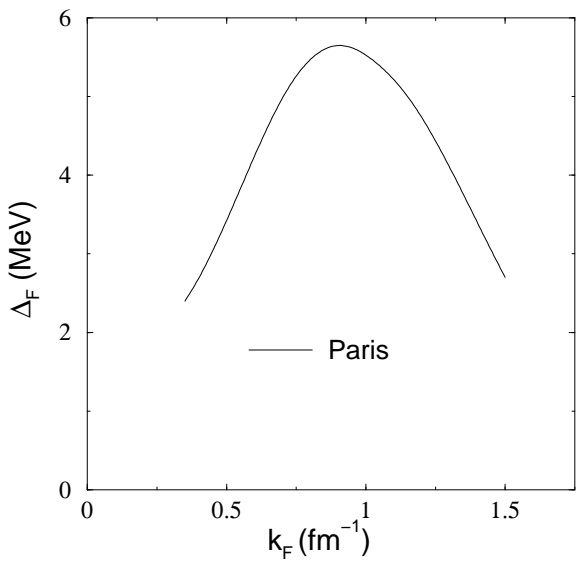,width=12.5cm,%
bbllx=3.0cm,bblly=-3cm,bburx=20cm,bbury=25.2cm,angle=270}}
\vspace{0.2cm}
\caption[]{
Pairing gap versus Fermi momentum for symmetric nuclear
matter in the $T=0$ channel from the Paris potential.
}
\end{figure}

The use of the bare force in the $T=0$ channel may, however, be more
questionable than in the $T=1$ channel. This stems from the implication of
the $d-$wave, i.e. the tensor force. The latter seems to be more affected by
medium effects than the $s-$wave part and therefore certainly great care
must be employed in this channel. In particular, it has been shown in \cite
{zamick} that higher shell admixtures make the tensor force appear weaker in
the valence space. Again the possible balance of the two graphs of Fig. 4
should thoroughly be investigated with respect to $s-$ and $d-$wave
contributions. We do not exclude the possibility that the tensor force is
largely screened in the medium and thus the enhancement of the $T=0$ gap
values may be brought back closer to values to which we are used in the $T=1$
case. However, without having detailed investigations at hand, we here stick
to our working hypothesis and base our considerations on the bare force
scenario. In this sense it may be interesting to also adjust, like we have
done it for the $T=1$ case, a density dependent $\delta -$force to the
calculation with the Paris force shown in Fig. 7. In principle, in this
case, the parameter $v_{0}$ should be chosen such that the deuteron binding
energy is reproduced in free space. We, however, found that with this
condition the cutoff parameter must be chosen very large rendering this
force not very practicable in actual calculations. We therefore adopted the
strategy to also vary within very narrow limits the parameter $v_{0}$ what
may slightly degrade the gap values at very low densities but significantly
improves them at the higher densities. In Fig. 8 we show such an adjustment
using various cutoffs. The value of $v_{0}$ used for the fits in Fig. 8 is 
\[
v_{0}=-1.05\;\frac{\hbar ^{2}}{m}\;\frac{2\pi ^{2}}{k_{C}}\;. 
\]
These fits should be useful for finite nuclei calculations.
\begin{figure}[t]
\centerline{\psfig{figure=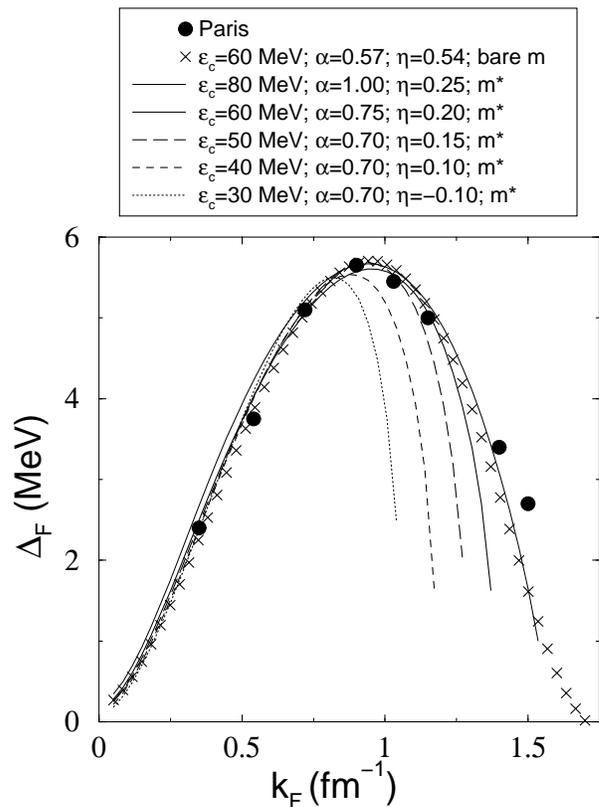,width=12cm,%
bbllx=-2cm,bblly=-2.6cm,bburx=20cm,bbury=21.cm,angle=0}}
\vspace*{-2cm}
\caption[]{
$T=0$ pairing gap in nuclear matter. The dots are the results
obtained from the Paris potential. The various curves correspond to fits
with Eq.(\ref{eqber}), using different parameters.
}
\end{figure}

\section{Concluding remarks}

In this paper we critically reviewed the use of effective nuclear pairing
forces. We argued that a Bethe-Goldstone or Brueckner G-matrix must not be
used in the gap equation. As a first guess, not knowing anything better, the
free nucleon-nucleon force may be tried in the gap equation. At least in the
traditional $T=1$ channel this prescription seems to work remarkably well,
since the best phenomenological force, namely the Gogny force, acts very
nearly like a free force in the $T=1$ pairing channel. We then advocated
that the same strategy should be adopted in the $T=0$ channel. We pointed
out that the situation may, however, be slightly more subtle there because
it is the action of the tensor force which makes the $T=0$ channel more
attractive than the $T=1$ one. The tensor part of the nuclear interaction
is, however, a very delicate subject and it may well be that it is more
affected by screening than the rest of the force. In the second part of the
work we demonstrated that the use of density dependent zero range forces in
the pairing channel may not be orthogonal to the use of finite range density
independent forces. Following Bertsch and Esbensen \cite{BE}, we give
parametrizations of density dependent $\delta -$forces which reproduce the
gap values in both $T=0$ and $T=1$ channels very well over the whole range
of relevant nuclear matter densities. Such forces, augmented by a cutoff,
should then also be useful for calculations in finite nuclei.

\acknowledgments 

We acknowledge useful discussions with N. Vinh Mau. This work was supported
by DGICYT-IN2P3 agreement and by DGICYT (Spain) under contract number
PB95/0123.

\end{multicols}

\end{document}